\documentclass[aps, twocolumn, pra, showpacs, superscriptaddress]{revtex4}
\usepackage{graphicx}
\usepackage{dcolumn}
\usepackage{bm}
\usepackage{amsmath}

\begin{document}
\title{Fidelity susceptibility in the two-dimensional transverse field Ising
and XXZ models}
\author{Wing-Chi Yu}
\affiliation{Department of Physics and ITP, The Chinese University of Hong Kong, Shatin,
Hong Kong, China}
\author{Ho-Man Kwok}
\affiliation{Department of Physics and ITP, The Chinese University of Hong Kong, Shatin,
Hong Kong, China}
\author{Junpeng Cao}
\affiliation{Department of Physics and ITP, The Chinese University of Hong Kong, Shatin,
Hong Kong, China}
\affiliation{Beijing National Laboratory for Condensed Matter Physics, Institute of
Physics, Chinese Academy of Sciences, Beijing 100080, China}
\author{Shi-Jian Gu}
\email{sjgu@phy.cuhk.edu.hk}
\affiliation{Department of Physics and ITP, The Chinese University of Hong Kong, Shatin,
Hong Kong, China}

\begin{abstract}
We study the fidelity susceptibility in the two-dimensional(2D)
transverse field Ising model and the 2D XXZ model numerically. It is
found that in both models, the fidelity susceptibility as a function
of the driving parameter diverges at the critical points. The
validity of the fidelity susceptibility to signal for the quantum
phase transition is thus verified in these two models. We also
compare the scaling behavior of the extremum of the fidelity
susceptibility to that of the second derivative of the ground state
energy. From those results, the theoretical argument that fidelity
susceptibility is a more sensitive seeker for a second order quantum
phase transition is also testified in the two models .
\end{abstract}

\date{\today}
\maketitle

%03.67.-a Quantum information (see also 42.50.Dv Quantum state
%engineering and measurements; 42.50.Ex Optical implementations of
%quantum information processing and transfer in quantum optics)

%64.70.Tg    Quantum phase transitions (for quantum Hall effects
%aspects, see 73.43.Nq in electronic structure of surfaces,
%interfaces, thin films, and low dimensional structures)

%75.10.-b General theory and models of magnetic ordering (see also
%05.50.+q Lattice theory and statistics)

%05.70.Fh    phase transitions: general studies

% 05.70.Jk Critical point phenomena

\section{Introduction}

Fidelity, a concept emerging from quantum information theory, has
recently become an attractive approach towards the study of critical
phenomena in condensed matter physics. In a quantum many-body
system, the quantum phase transition is completely driven by the
quantum fluctuation in the ground state and is incarnated by an
abrupt change in the qualitative structure of the ground state
wavefunction as the system varies across the critical point
\cite{Sachdev}. Therefore, being a measure of the similarity between
two states, the fidelity is excepted to show a dramatic change
across the transition points. This motivated people to start
exploring its role played in quantum phase transitions
\cite{HTQuan,PZana,HQZhou0701}. Moreover, as the fidelity can be
viewed as a space geometrical quantity, no a priori knowledge of the
order parameter and symmetry breaking of the system is required.
This is thus a great advantage to the study of quantum phase
transitions using fidelity approaches.

%Unfortunately, the calculation of fidelity from the ground state
%wavefunction is usually quite involving and it depends on the small but yet
%finite driving parameter that enters the Hamiltonian.

Following the streamline of fidelity, some alternative schemes, like
the fidelity susceptibility \cite{WLYou}, fidelity per site
\cite{HQZhou}, operator fidelity \cite{XWang}, and
density-functional fidelity \cite{SJGu}, have been proposed. As to
establish a closer picture to condensed matter physics, we follow
the concept of fidelity susceptibility in this paper.
Mathematically, the fidelity susceptibility is just the leading term
of the fidelity. It defines the response of the fidelity to the
driving parameter. As a result, the singularity of the fidelity
across the transition points could thus be reflected in the
divergence of the fidelity susceptibility. In fact, this argument
has been consolidated by the results in a number of one-dimensional
quantum many-body systems \cite{fs1D} (See also a review article
\cite{GuRev}).

In this paper, we investigate the behavior of the fidelity
susceptibility in two two-dimensional (2D) models, namely the 2D
transverse field Ising model and the XXZ model numerically. Our
results show that the fidelity susceptibility as a function of the
driving parameter diverges at the quantum phase transition points in
both of the models. The scaling behavior of the extremum of the
fidelity susceptibility at the transition point and that of the
second derivative of the ground state energy are also compared. From
those results, the theoretical argument that the fidelity
susceptibility is a more sensitive indicator than the second
derivative of the ground state energy in searching for a second
order quantum phase transition is testified. Besides, it is also
found that the fidelity susceptibility shows a scaling behavior in
the vicinity of the critical point and its critical exponents for
both models are also obtained through finite-size scaling analysis.

\section{Formulism}

For a general form of the Hamiltonian,
\begin{equation}
H(\lambda )=H_{0}+\lambda H_{I},
\end{equation}%
where $H_{I}$ is the driving Hamiltonian and $\lambda $ denotes its
strength. The fidelity is the modulus of the overlap between two ground
states $|\Psi _{0}(\lambda )\rangle $ and $|\Psi _{0}(\lambda +\delta
\lambda )\rangle $ \cite{PZana},
\begin{equation}
F(\lambda ,\lambda +\delta \lambda )=|\langle \Psi _{0}(\lambda )|\Psi
_{0}(\lambda +\delta \lambda )\rangle |.
\end{equation}%
Since our focus is on continuous quantum phase transitions, the
ground state of the Hamiltonian is non-degenerated for a finite
system. $|\Psi _{0}(\lambda +\delta \lambda )\rangle $ can thus be
obtained from the time-independent non-degenerated perturbation
theory. Extracting the leading term of the fidelity, the fidelity
susceptibility can be expressed as \cite{WLYou,PGio}
\begin{equation}
\chi _{F}(\lambda )=\sum_{n\neq 0}\frac{|\langle \Psi _{n}(\lambda
)|H_{I}|\Psi _{0}(\lambda )\rangle |^{2}}{[E_{n}(\lambda )-E_{0}(\lambda
)]^{2}},  \label{eq:fspert}
\end{equation}%
where $|\Psi _{n}(\lambda )\rangle $ is a set of orthogonal basis satisfying
$H(\lambda )|\Psi _{n}(\lambda )\rangle =E_{n}(\lambda )|\Psi _{n}(\lambda
)\rangle $. On the other hand, consider the second derivative of the ground
state energy with respect to $\lambda $ \cite{SChen},
\begin{equation}
\frac{\partial ^{2}E_{0}(\lambda )}{\partial \lambda ^{2}}=\sum_{n\neq 0}%
\frac{2|\langle \Psi _{n}(\lambda )|H_{I}|\Psi _{0}(\lambda )\rangle |^{2}}{%
E_{0}(\lambda )-E_{n}(\lambda )},
\end{equation}%
one may easily realize that the above expression is very similar to the
perturbation form of fidelity susceptibility in Eq. (\ref{eq:fspert}) except
having different exponent in the denominator. Therefore, one may expect that
both the singularity of fidelity susceptibility and the second derivative of
the ground state energy are intrinsically due to the vanishing of the energy
gap in the thermodynamic limit \cite{SChen}. However, the difference in the
exponent of the dominator makes fidelity susceptibility a more sensitive
quantity in searching for quantum phase transitions. That is to say, while
the fidelity susceptibility shows a divergence at the critical point, the
second derivative of the ground state energy may still be a continuous
function.

\begin{figure}[tbp]
\includegraphics[width=8cm]{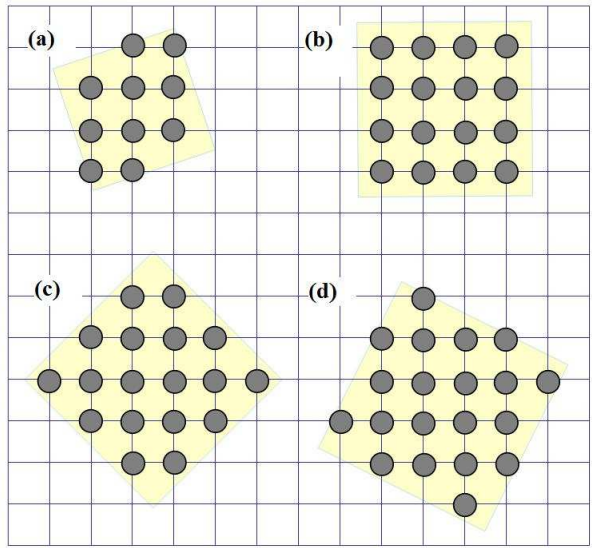}
\caption{(Color online) Two-dimensional structures for different system sizes
$N=10, 16, 18, 20$ which can be placed on a square lattice with periodic
boundary conditions.} \label{fig:2dconfig}
\end{figure}

Furthermore, to study the scaling behavior of the fidelity
susceptibility around
the critical point, we may perform finite-size scaling analysis \cite%
{LCVenuti07,Guscal}. Let's consider a system consisting of $N$ sites such
that $N=L^d$, where $d$ is the real dimension of the system. Around the
critical point $\lambda_c$, the fidelity susceptibility behaves as
\begin{eqnarray}
\frac{\chi_{F}(\lambda)}{L^{d_a^\pm}}\sim\frac{1}{|\lambda_c-\lambda|^{%
\alpha^\pm}},
\end{eqnarray}
where $\alpha^+$($\alpha^-$) is the critical exponent of the fidelity
susceptibility above (below) the critical point, $d_a^\pm$ is the quantum
adiabatic dimension and hence $\chi_{F}(\lambda)/L^{d_a^\pm}$ is an intensive
quantity. For a finite system, if the fidelity susceptibility shows a peak at a
certain point $\lambda_{\rm max}$, it's maximum value scales like
\begin{eqnarray}
\chi_F(\lambda_{\mathrm{max}})\sim L^{d_a^c},
\end{eqnarray}
where $d_a^c$ is the critical adiabatic dimension. The above two asymptotic
behaviors satisfy \cite{Guscal}
\begin{eqnarray}
\frac{\chi_F(\lambda,L)}{L^{d_a^\pm}} = \frac{A}{L^{-d_a^c+d_a^\pm}+ B
(\lambda-\lambda_{\mathrm{max}})^{\alpha^\pm}},  \label{eq:scafun}
\end{eqnarray}
where $A$ is a constant, B is a non-zero function of $\lambda$ and both of them
are independent of the system size. From Eq. (\ref{eq:scafun}), one can
find that the rescaled fidelity susceptibility is an universal function of $%
L^\nu(\lambda-\lambda_{\mathrm{max}})$,
\begin{eqnarray}
\frac{\chi_F(\lambda_{\mathrm{max}},L)-\chi_F(\lambda,L)}{\chi_F(\lambda,L)}%
=f[L^\nu (\lambda-\lambda_{\mathrm{max}})],
\end{eqnarray}
where $\nu$ is the critical exponent of the correlation length. The critical
exponent of the fidelity susceptibility can then be obtained as \cite%
{Guscal,SJGuqad}
\begin{eqnarray}
\alpha^\pm=\frac{d^c_a-d_a^{\pm}}{\nu}.  \label{eq:exp}
\end{eqnarray}

\begin{figure}[tbp]
\includegraphics[width=8cm]{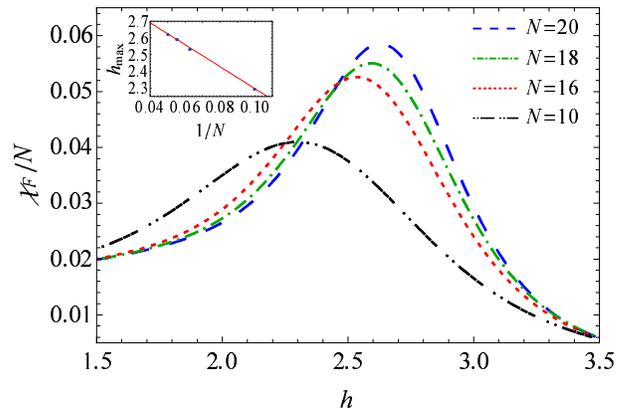}
\caption{(Color online)  The averaged fidelity susceptibility in the ground
state of the 2D transverse field Ising model on a square lattice as a function
of $h$. The inset shows $h_{\mathrm{max}}$ as a function of $1/N$. The
$y$-intercept of the line is $2.95\pm 0.01$. } \label{fig:ising_fs_h}
\end{figure}

\begin{figure}[tbp]
\includegraphics[width=8cm]{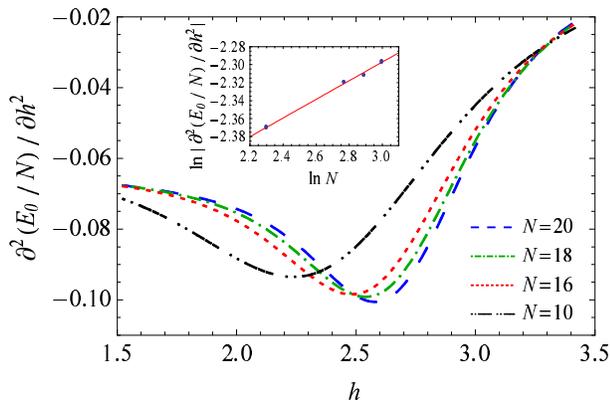}
\caption{(Color online)  The second derivative of the averaged ground state
energy of the 2D transverse field Ising model on a square lattice as a function
of $h$. The inset shows the scaling behavior of the minimum of the second
derivative
of the averaged ground state energy. The slope of the line is approximately $%
0.103$.}
\label{fig:ising_gse}
\end{figure}

\section{Two-dimensional Transverse Field Ising Model}

The Hamiltonian of the 2D transverse field Ising model
\cite{deGennes1963,Stinchombe1973} defined on a square lattice reads
\begin{equation}
H_{\text{Ising}}=-\sum_{\langle \mathbf{ij}\rangle }S_{\mathbf{i}}^{x}S_{%
\mathbf{j}}^{x}-\frac{h}{2}\sum_{\mathbf{i}}S_{\mathbf{i}}^{z},
\end{equation}%
where $S_{\mathbf{i}}^{x}$, $S_{\mathbf{i}}^{y}$, $S_{\mathbf{i}}^{z}(S_{%
\mathbf{i}}^{\kappa }=\sigma _{\mathbf{i}}^{\kappa }/2,\kappa =x,y,z)$ are
spin-$1/2$ operators at site \textbf{i}, $h$ is the transverse field strength
in unit of the Ising coupling, and the sum $\langle \rangle $ runs over the
nearest-neighboring pairs on the lattice. Periodic boundary conditions are
assumed. This model was originally introduced by de Gennes to describe
potassium-dihydrogen-phosphate type ferroelectrics \cite{deGennes1963} and has
been studied extensively via various approaches, like real-space
renormalization group \cite{Friedman1978}, density-matrix renormalization
\cite{deJongh1998}, numerical diagonalization\cite{MHenkel,CJHamer}, and
entanglement \cite{Olav2004}.

Obviously, the Hamiltonian of the model commutes with the parity
operator $P=\prod_{\mathbf{i}}\sigma _{\mathbf{i} }^{z}$. Thus each
eigenstate of the Hamiltonian is also an eigenstate of $P$. The
Hilbert space can then be decomposed into subspace $V(p)$ where $p$
is the eigenvalue of $P$ and is specified in each subspace. For a
finite system, the ground state of the Ising model is non-degenerate
in each subspace, thus the perturbation expansion as introduced in
the previous section is valid as long as the lattice is finite. In
the thermodynamic limit, the model exhibits a quantum phase
transition at $1/h_{c}\simeq 0.328$ \cite{MHenkel,CJHamer}. For
$h\gg h_{c}$, the transverse field dominates and the ground state is
a paramagnetic phase, with spins almost fully polarized in the
$z$-direction. For $h\ll h_{c}$, the ground state is a ferromagnetic
phase and is doubly degenerated.

To study a model on a two-dimensional square lattice with periodic
boundary conditions, we need to construct proper lattice structures
that are suitable for exact diagonalization. In this paper, we will
diagonalize two models with
system sizes $N=10, 16, 18, 20$, whose structures are shown in Fig. \ref%
{fig:2dconfig}. The effective length $L=\sqrt{N}$ might then be a
real number instead of an integer.

Fig. \ref{fig:ising_fs_h} shows the numerical result of the fidelity
susceptibility of the Ising model on a square lattice for various
system sizes. It can be seen that on both sides around the critical
point, the averaged fidelity susceptibility is an intensive
quantity, i.e. $\chi _{F}\sim N\sim L^{2}$ and we have $d_{a}^{\pm
}=2$. More importantly, the averaged fidelity susceptibility for
different $N$ all show a peak at $h_{\mathrm{max}}$. This peak
position of the fidelity susceptibility $h_{\mathrm{max}}$ is
plotted as a function of $1/N$, as shown in the inset of Fig.
\ref{fig:ising_fs_h}. The linear fitting gives
\begin{equation*}
h_{\mathrm{max}}=2.95-\frac{6.56}{N}.
\end{equation*}%
In the thermodynamic limit, we obtain
\begin{equation*}
h_{c}=2.95\pm 0.01,
\end{equation*}%
which gives $1/h_{c}=0.326\pm 0.001$. Comparing this value to the critical point $%
1/h_{c}\simeq 0.328$ obtained in previous works
\cite{MHenkel,CJHamer}, our result here is consistent with them up
to two digits.

Moreover, we can also see that the averaged fidelity susceptibility
peaks sharper for a larger $N$ and is in fact scales
approximately with $N^{0.51}$ as shown in the inset of Fig. \ref%
{fig:ising_scaling}. Physically, as the ground state wavefunction of
the model changes abruptly across the transition point, the fidelity
susceptibility, as a measure of the leading response of the fidelity
to the driving parameter, is intuitively expected to show a
divergence at the critical point. Here, we have shown numerically
this is in fact the case and verified the significance of fidelity
susceptibility in signaling for the quantum phase transition in the
2D Ising model.

Fig. \ref{fig:ising_gse} shows the second derivative of the averaged
ground state energy of the Ising model for several system size of a
square lattice as a function of $h$. As it is well-known that the
Ising model exhibits a second order phase transition, the second
derivative of the averaged ground state energy is expected to show a
minimum at the transition point. From the inset of Fig.
\ref{fig:ising_gse}, it is found that the minimum value of the
second derivative of the averaged ground state energy scales
approximately with $N^{0.103}$. Comparing this value with that of
the fidelity susceptibility, which is about $0.51(d_{a}^{c}\simeq
3.02)$, we may conclude that the fidelity susceptibility is a more
sensitive tool in detecting for a second order quantum phase
transition.

\begin{figure}[tbp]
\includegraphics[width=8cm]{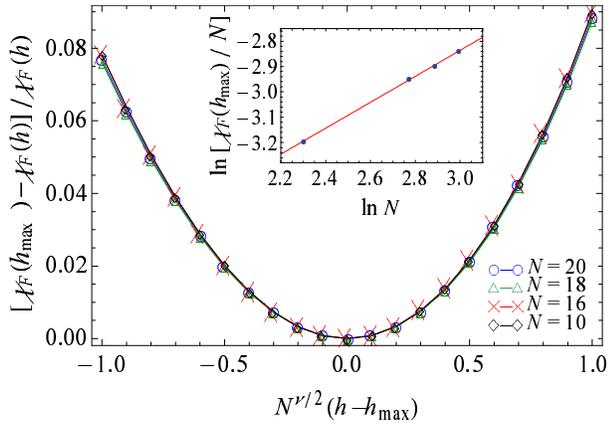}
\caption{(Color online) The finite-size scaling analysis is performed for the
case of power-law divergence for the 2D transverse field Ising model. The
fidelity
susceptibility for different system sizes is a function of $N^{\protect\nu%
/2}( h-h_{\mathrm{max}})$ only, with the critical exponent $\protect\nu\simeq
1.40$. The insert shows the scaling behavior of the maximum of the averaged
fidelity susceptibility. The slope of the line is 0.51.}
\label{fig:ising_scaling}
\end{figure}

Fig. \ref{fig:ising_scaling} shows the finite-size scaling analysis in the
case of power-law divergence of the 2D transverse field Ising model. The
rescaled fidelity susceptibility collapsed to a single curve for various
system sizes. The critical exponent of the correlation length can thus be
obtained as $\nu \simeq 1.40$. Together with the slope of the line in the
inset of Fig. \ref{fig:ising_scaling} and from Eq. (\ref{eq:exp}), the
critical exponent of the fidelity susceptibility is found to be
\begin{equation*}
\alpha =\frac{1.02}{1.40}\simeq 0.73.
\end{equation*}

\begin{figure}[tbp]
\includegraphics[width=8cm]{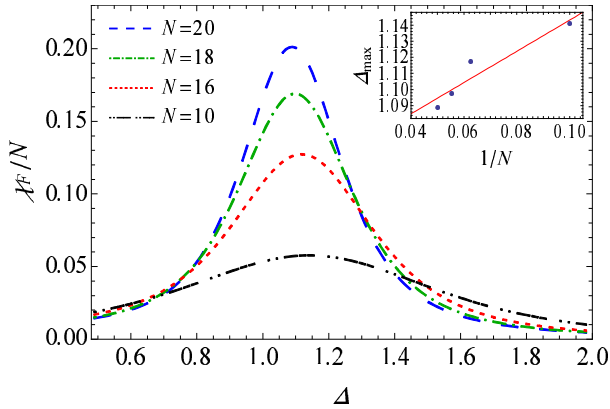}
\caption{(Color online) The averaged fidelity susceptibility in the ground
state of the 2D
XXZ model on a square lattice as a function of $\Delta$. The inset shows $%
\Delta_{\mathrm{max}}$ as a function of $1/N$. The $y$-intercept of the line
is $1.05\pm 0.02$.}
\label{fig:XXZ_fs}
\end{figure}

\begin{figure}[tbp]
\includegraphics[width=8cm]{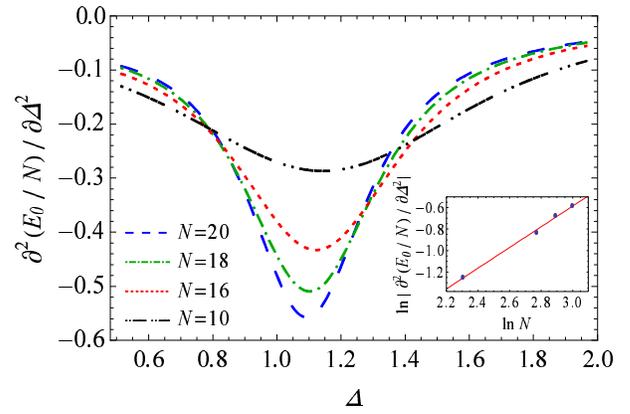}
\caption{(Color online) The second derivative of the averaged ground state
energy of the 2D XXZ model as a function of $\Delta$. The inset shows the
scaling behavior of the minimum of the second derivative of the averaged ground
state energy. The slope of the line is approximately $0.96$.}
\label{fig:XXZ_gse}
\end{figure}

\begin{figure}[tbp]
\includegraphics[width=8cm]{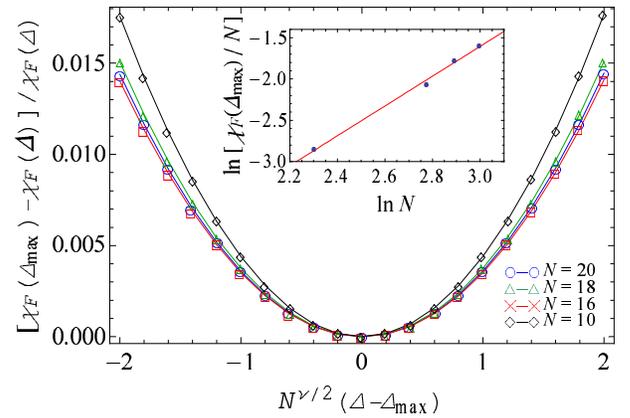}
\caption{(Color online) The finite-size scaling analysis is performed for the
case of power-law divergence for the 2D XXZ model. The fidelity susceptibility
for $N>10$ is a function of
$N^{\protect\nu/2}(\protect\Delta-\protect\Delta_{\mathrm{max}})$ only, with
the critical exponent $\protect\nu\simeq 3.00$. The insert shows the scaling
behavior of the maximum of the averaged fidelity susceptibility. The slope of
the line is 1.81.} \label{fig:XXZ_scaling}
\end{figure}

\section{Two-dimensional XXZ Model}

For the 2D XXZ model, the Hamiltonian is given by
\begin{equation}
H_{\text{XXZ}}=\sum_{\langle \mathbf{ij}\rangle }(S_{\mathbf{i}}^{x}S_{%
\mathbf{j}}^{x}+S_{\mathbf{i}}^{y}S_{\mathbf{j}}^{y}+\Delta S_{\mathbf{i}%
}^{z}S_{\mathbf{j}}^{z}),
\label{eq:H_XXZ}
\end{equation}%
where $\Delta =J_{z}/J_{y}(J_{x}=J_{y})$ is the dimensionless
parameter characterizing the anisotropy of the model. The sum is
over all nearest-neighbors on a square lattice. Again, periodic
boundary conditions are assumed. For the XXZ model in two
dimensions, there exists no exact solution. One has to use either
approximate analytical approach such as the spin-wave theory or
numerical approach such as exact diagonalization studies of a finite
lattice. For the latter approach, to obtain results in the
thermodynamic limit, finite-size scaling analysis must be performed
\cite{Sandvik1997,HQLin2001}. Therefore, a physical quantity which
is more sensitive to the system size than the traditional second
derivative of the ground state energy would be very useful to study
the critical phenomena numerically.

From Eq. (\ref{eq:H_XXZ}), it can be easily seen that the
Hamiltonian of the XXZ model commutes with the $z$-component of
total spin operator
$S_{\mathrm{total}}^{z}=\sum_{\mathbf{i}}S_{\mathbf{i}}^{z}$.
Thus, each eigenstate of the Hamiltonian is also an egienstate of $S_{%
\mathrm{total}}^{z}$. The Hilbert space can then be decomposed into
numerous subspaces $V(M)$, where $M$ is the eigenvalue of
$S_{\mathrm{total}}^{z}$. For a finite sample, the ground state of
the XXZ model is non-degenerated in any of the admissible subspace
$V(M)$ \cite{ELieb,IAffleck}. Therefore, the perturbation expansion
can also be applied to this model as long as the system is finite.
In the thermodynamic limit, the quantum phase transition takes place
at the isotropic point $\Delta_{c}=1$. This phenomenon can be
understood by the picture of the first excited energy levels
crossing at the transition point \cite{GSTian2003}. For $\Delta \gg
\Delta _{c}$, the last term in the Hamiltonian dominates and the
ground state is an antiferromagnetic phase along the $z$-direction.
For $\Delta \ll \Delta _{c}$, the first two terms in the Hamiltonian
dominate and the ground state is also an antiferromagnetic phase,
but in the $xy$ plane. It is well known that long-range orders are
present in both of the two phases. However, whether there's a
long-range order at the critical point is still an open question.
With the help of fidelity susceptibility, we may also find some
hints towards this question.

The numerical result of the averaged fidelity susceptibility for
various system sizes of the 2D XXZ model on a square lattice as a
function of $\Delta $ is shown in Fig. \ref{fig:XXZ_fs}. The
averaged fidelity susceptibility is an intensive quantity, meaning
that $\chi _{F}\sim N$, on both sides of the critical point.
Moreover, like the previous case of the Ising model, the
averaged fidelity susceptibility of the XXZ model also shows a peak at $%
\Delta _{\mathrm{max}}$. The inset of Fig. \ref{fig:XXZ_fs} shows the peak
position of the fidelity susceptibility $\Delta _{\mathrm{max}}$ as a
function of $1/N$. The linear fitting gives
\begin{equation*}
\Delta _{\mathrm{max}}=1.05+\frac{0.97}{N}.
\end{equation*}%
In the thermodynamic limit, we obtain
\begin{equation*}
\Delta _{c}=1.05\pm 0.02.
\end{equation*}%
Comparing to the theoretical critical point $\Delta _{c}=1$, our
result here is consistent up to two digits.

Besides, from the slope of the straight line in the inset of Fig.
\ref{fig:XXZ_scaling}, it is found that the peak of the averaged
fidelity susceptibility scales with the system size like $N^{1.81}$.
Therefore, one may expect the fidelity susceptibility to show a
singularity at the critical point in the thermodynamic limit. Hence,
the validity of the fidelity susceptibility as a seeker for the
quantum phase transition is also verified in the 2D XXZ model.
Nevertheless, following the idea of the implication of existence of
long-range correlation from the divergence of the fidelity
susceptibility \cite{GuLRO}, we argue that long-range correlation is
in fact present at the transition point of the 2D XXZ model. This is
also in agreement with the
previous conclusion drawn from the study of the 2D XXZ model using entanglement \cite%
{GuXXZentan}.

In comparison, the second derivative of the averaged ground state
energy for various system sizes exhibits a minimum at the transition
point, as shown in Fig. \ref{fig:XXZ_gse}. From the inset of Fig.
\ref{fig:XXZ_gse}, it is also found that the minimum value of the
second derivative of the averaged ground state energy scales
approximately with $N^{0.96}$, meaning that it shows a slower
divergence at the critical point compared to the fidelity
susceptibility. In other words, the fidelity susceptibility is again
a more sensitive candidate in seeking for the quantum phase
transition in the 2D XXZ model.

Fig. \ref{fig:XXZ_scaling} shows the finite-size scaling analysis in the
case of power-law divergence of the 2D XXZ model. The rescaled fidelity
susceptibility almost collapsed to a single curve for a large enough system
size, say $N>10$. The exponent of the correlation lenght is obtained as $%
\nu\simeq 3.00$. From the slope of the inset in Fig.
\ref{fig:XXZ_scaling} and using Eq. (\ref{eq:exp}), the critical
exponent of the fidelity susceptibility is calculated to be
\begin{eqnarray*}
\alpha=\frac{3.62}{3.00}=1.21.
\end{eqnarray*}

\section{Summary}

To conclude, through the numerical study of the fidelity susceptibility in the
2D transverse field Ising model and the 2D XXZ model, we found that the
fidelity susceptibility as a function of the driving parameter diverges in both
models at the critical point. By comparing the scaling behavior of the extremum
of the fidelity susceptibility to that of the second derivative of the ground
state energy, we also showed that fidelity susceptibility is a more sensitive
indicator in detecting for a second order quantum phase transition. By
performing finite-size scaling analysis, the critical exponent of the fidelity
susceptibility in both models are also obtained. Finally, due to the divergence
of fidelity susceptibility in the 2D XXZ model, we argued that the system shows
a long-range correlation at the critical point of the model.

This work is supported by the Earmarked Grant for Research from the Research
Grants Council of HKSAR, China (Project No. CUHK 400807).

\end{document}